\documentclass{aastex631}
\usepackage{threeparttable} 
\usepackage{graphicx} 
\usepackage{rotating} 
\usepackage{amsmath}
\usepackage{booktabs}
\usepackage[utf8]{inputenc}
\usepackage{booktabs}
\usepackage{multirow}
\usepackage{xcolor}
\usepackage{threeparttable}

\defcitealias{Kajikiya2024}{Paper~I}
\begin{document}

\title{High-Time-Cadence Spectroscopy and Photometry of Stellar Flares on M-dwarf YZ Canis Minoris with Seimei Telescope and TESS. II. Statistical Properties of Blue/Red Asymmetries in the H$\alpha$ Line}

\author[0009-0000-2704-9785]{Yuto Kajikiya}
\affiliation{Department of Earth and Planetary Sciences, Institute of Science Tokyo \\
2-12-1 Ookayama, Meguro-ku, Tokyo 152-8551, Japan}

\author[0000-0002-1297-9485]{Kosuke Namekata}
\affiliation{Heliophysics Science Division, NASA Goddard Space Flight Center, 8800 Greenbelt Road, Greenbelt, MD 20771, USA}
\affiliation{The Catholic University of America, 620 Michigan Avenue, N.E. Washington, DC 20064, USA}
\affiliation{The Hakubi Center for Advanced Research, Kyoto University, Kyoto 606-8302, Japan}
\affiliation{Department of Physics, Kyoto University, Kitashirakawa-Oiwake-cho, Sakyo-ku, Kyoto, 606-8502, Japan}
\affiliation{Division of Science, National Astronomical Observatory of Japan, NINS, Osawa, Mitaka, Tokyo, 181-8588, Japan}

\author[0000-0002-0412-0849]{Yuta Notsu}
\affiliation{Laboratory for Atmospheric and Space Physics, University of Colorado Boulder, 3665 Discovery Drive, Boulder, CO 80303, USA}
\affiliation{National Solar Observatory, 3665 Discovery Drive, Boulder, CO 80303, USA}

\author[0000-0002-5978-057X]{Kai Ikuta}
\affiliation{Department of Multidisciplinary Sciences, The University of Tokyo, 3-8-1 Komaba, Meguro, Tokyo 153-8902, Japan}

\author[0000-0003-0332-0811]{Hiroyuki Maehara}
\affiliation{Okayama Branch Office, National Astronomical Observatory of Japan\\ 
3037-5 Honjo, Kamogata-cho, Asakuchi, Okayama 719-0232, Japan}

\author[0000-0001-8033-5633]{Bunei Sato}
\affiliation{Department of Earth and Planetary Sciences, Institute of Science Tokyo \\
2-12-1 Ookayama, Meguro-ku, Tokyo 152-8551, Japan}

\author[0000-0001-9588-1872]{Daisaku Nogami}
\affiliation{Department of Astronomy, Kyoto University, 
Kitashirakawa Oiwake-cho, Sakyo-ku, Kyoto 606-8502, Japan}

\begin{abstract}
M-dwarfs frequently produce flares, and their associated coronal mass ejections (CMEs) may threaten the habitability of close-in exoplanets. M-dwarf flares sometimes show prominence eruption signatures, observed as blue/red asymmetries in the H$\alpha$ line. In Paper I, we reported four candidates of prominence eruptions, which shows large diversity in their durations and velocities. In this study, we statistically investigate how blue/red asymmetries are related with their flare and starspot properties, using the dataset from 27 H$\alpha$ flares in Paper I and previously reported 8 H$\alpha$ flares on an M-dwarf YZ Canis Minoris. We found that these asymmetry events tend to show larger H$\alpha$ flare energies compared to non-asymmetry events. In particular, 5 out of 6 blue asymmetry events are not associated with white-light flares, whereas all 7 red asymmetry events are associated with white-light flares. Furthermore, their starspot distributions estimated from the TESS light curve show that all prominence eruption candidates occurred when starspots were located on the stellar disk center as well as on the stellar limb. These results suggest that flares with lower heating rates may have a higher association rate with prominence eruptions and/or the possibility that prominence eruptions are more detectable on the limb than on the disk center on M-dwarfs. These results provide significant insights into CMEs that can affect the habitable world around M-dwarfs.
\end{abstract}


\section{Introduction} \label{sec:intro}
Solar flares are the most energetic phenomena on the Sun \citep{Shibata&Magara2011} and are often accompanied by coronal mass ejections (CMEs) (e.g., \citealt{Cliver2022}), which can significantly impact Earth's environment (e.g., \citealt{Bisi2010}; \citealt{Gopalswamy2016}). Recent stellar observations have reported that active young M-dwarfs produce flares that are both more frequent and more energetic than those of the Sun. Additionally, superflares with energies exceeding ten times that of the largest solar flares are also frequently observed on young M-dwarfs (e.g., \citealt{Kowalski2024}). On the other hand, M-dwarfs are among the most promising targets in the search for life beyond the Solar System (e.g., \citealt{Nutzman2008,Gilbert2020}). Their habitable zones are located close to the host star (e.g., \citealt{Kasting2014}). However, large CMEs associated with strong flares could significantly impact these close-in habitable planets. They may strip away planetary envelopes, leading to atmospheric escape \citep{Lammer2007}. On the other hand, high-energy particles accelerated by CMEs could trigger nitrogen fixation and drive prebiotic chemistry. This could potentially lead to the production of greenhouse gases as well as the formation of complex organic molecules, such as amino acids \citep{Airapetian2016,Kobayashi2023}. Moreover, frequent CMEs could solve the UV paucity problem for prebiotic chemistry around M-dwarfs \citep{Ranjan2017,Lingam2018}.

Previous optical spectroscopic observations of M-dwarf flares have reported that chromospheric line profiles often show blue/red asymmetries (e.g., \citealt{Houdebine1990,Vida2016,Vida2019,Maehara2021,Notsu2024,Inoue2024,Kajikiya2024}). The red asymmetric components may suggest chromospheric condensation or post flare loop \citep{Namizaki2023}, while the blue asymmetric components are though to indicate prominence eruptions, which form the core of CMEs, or other flare-related phenomena (e.g., \citealt{Maehara2021,Notsu2024}). A few of these eruption candidates show velocities exceeding the escape velocity of the star \citep{Houdebine1990,Vida2016}, which might serve as indirect evidence of CMEs. On the other hand, their typical velocities are around $\sim$100 km s$^{-1}$, which is much slower than the surface escape velocity of M-dwarfs ($\sim$600 km s$^{-1}$). The kinetic energies of these eruption candidates are 1--2 orders of magnitude lower than those of the extrapolated values from the scaling relations of solar CME velocities and bolometric flare energy \citep{Moschou2019,Maehara2021,Notsu2024}. Moreover, the association rate of flares with prominence eruption candidates is relatively low on M-dwarfs ($\sim$20\%; \citealt{Notsu2024,Kajikiya2024}) compared to the higher rate observed in the active G-dwarf EK Dra ($\sim$60\%; \citealt{Namekata2024a}), although the sample size is not enough for statistics. In the case of the Sun, more than 90\% of X-class flares are accompanied by CMEs \citep{Yashiro&Gopalswamy2009}. Most events observed in the H$\alpha$ line on M-dwarfs have bolometric energies exceeding $10^{32}$ erg, corresponding to X10-class solar flares. Therefore, the low association rate in M-dwarfs contradicts solar observations. (e.g., \citealt{Notsu2024,Kajikiya2024}). This low association rate of prominence eruption candidates on M-dwarfs might be due to differences in the underlying physical mechanisms or observational biases \citep{Odert2020}. 
Magnetohydrodynamic simulations assuming large-scale strong dipole magnetic fields suggest that strong overlying coronal magnetic fields may suppress such eruptions \citep{Drake2016,Alvarado-Gomez2018,Sun2022}. Zeeman Doppler Imaging observations suggest that M-dwarfs have strong overlying magnetic fields (e.g., \citealt{Morin2008}). Therefore, this magnetic suppression mechanism is one possible hypothesis to explain the low association rate, as well as the low velocities and kinetic energy, of prominence eruptions on M-dwarfs (e.g., \citealt{Maehara2021,Notsu2024}). To investigate the potential role of this magnetic suppression and possibility of observational biases, it is necessary to statistically examine the dependence of prominence eruption candidates on flare properties (e.g., energy, duration) and starspot properties.

One-month continuous photometric observations by the Transiting Exoplanet Survey Satellite (TESS, \citealt{2015JATIS...1a4003R}) have enabled long-term simultaneous photometric and spectroscopic monitoring observations. Photometric observations, simulutenously with spectroscopic observations, can characterize white-light-flares and starspot properties, which could be helpful to infer the origin of the blue asymmetry. First, white-light flares are thought to represent emissions mainly from flare footpoints and originate from non-thermal electrons \citep{Watanabe2017,Namekata2017,Namizaki2023}. In the case of the Sun, impulsive flares tend to show more significant white-light emission in flare ribbons than long-duration flares \citep{Watanabe2017}, so the presence of white-light emission indicates the impulsive flare heating. Additionally, white-light emission would not be visible when flare footpoints are obscured near the limb. 
Therefore, simultaneous photometric observations contribute not only to a better understanding of the flare heating properties but also to constraining the locations of flare origins.
Second, photometric observations can capture the rotational modulation caused by starspots, which also enables the estimation of starspot distributions (e.g., \citealt{Ikuta2023}). Therefore, simultaneous observations with TESS give us a good opportunity to investigate statistical relationships between prominence eruptions, flare properties and spot properties. However, the number of such observations is small, and no statistical studies have been conducted using simultaneous observation data with TESS.


\cite{Kajikiya2024} \citepalias[hereafter,][]{Kajikiya2024} conducted optical spectroscopic observations of magnetically active M4.5-type dwarf YZ CMi using the 3.8m Seimei Telescope \citep{Kurita2020} simultaneously with TESS from January 18 to February 4, 2021. Our spectroscopic observations were conducted with an unprecedentedly short time cadence of  $\sim$1 minute.
We detected 27 H$\alpha$ flares, as a result of our long term observations.
Among them, 5 flares show red asymmetries, and 3 flares show blue asymmetries in the H$\alpha$ line profile. 
We focused on the events showing blue and red asymmetries and discussed, for each individual event, whether they are likely to be prominence eruptions.
A total of 4 events (3 blue asymmetries and 1 red asymmetry) suggest prominence eruptions based on their velocities and time evolution. This increased the sample size of prominence eruption candidates for statistical analysis.

In this study, we investigate the statistical dependency of prominence eruption candidates as well as other asymmetry events on flare properties and starspot properties. 
In Section \ref{sec:Observation and Data}, we summarize the observational dataset used in this study. In Section \ref{sec:Asymmetries vs Flare properties}, we examine the dependence of asymmetry types on flare properties (energy and duration in Section 3.1, and white-light enhancement in Section 3.2). In Section \ref{sec:Asymmetries vs Spot Properties}, we investigate the dependence of asymmetry types on starspot properties. The summary and conclusions are provided in Section \ref{sec:Summary and Conclusions}.

\section{Observation Data} \label{sec:Observation and Data}
Here, we summarize the dataset used in this work. We primarily use the data of all the 27 H$\alpha$ flares on an active M-dwarf YZ CMi reported in \citetalias{Kajikiya2024}. In addition to \citetalias{Kajikiya2024} dataset, we also incorporate the data of flares on the same M-dwarf YZ CMi from the two previous studies \citep{Maehara2021,Notsu2024}, which conducted simultaneous observations with TESS photometry and optical spectroscopy like \citetalias{Kajikiya2024}. \citet{Maehara2021} used the 2.0 m Nayuta telescope at the Nishi-Harima Astronomical Observatory for spectroscopy, while \citet{Notsu2024} used the ARC 3.5 m telescope at Apache Point Observatory (APO) for spectroscopy. Table \ref{tab:Flares_spot_discussion} lists the observational properties of flares used in this paper.
\cite{Maehara2021} conducted spectroscopic observations simultaneously with TESS Cycle 1 and detected 4 H$\alpha$ flares. They discussed line asymmetry for only two events, one of which showed blue asymmetry. Therefore, we incorporate only these two events (M1 and M2 in Table \ref{tab:Flares_spot_discussion}).
\cite{Notsu2024} conducted simultaneous optical spectroscopic and photometric (TESS Cycle 1,3 or ground-based) observations of mid-M dwarfs YZ CMi, EV Lac and AD Leo during the 31 nights over 2 yr (2019 January–2021 February). They detected 41 H$\alpha$ flares. Among these, five (N1--N5 in Table \ref{tab:Flares_spot_discussion}) were observed simultaneously with TESS Cycle 1 and Cycle 3 on YZ CMi, of which one showed a red asymmetry and another showed a blue asymmetry. We note that some H$\alpha$ flares (Y16 and N2) show two independent red and blue asymmetries, respectively. Then, the number of H$\alpha$ flares showing asymmetry differs from the number of asymmetry events. Table \ref{tab:asymmetry_events} lists the number of asymmetry events in different studies. 

\begin{table}[ht]
\centering
\resizebox{\textwidth}{!}{%
\begin{threeparttable}
\caption{Observational properties of flares on YZ CMi used in this paper}
\label{tab:Flares_spot_discussion}
\begin{tabular}{ccccccccccc}
\hline\hline
ID\tnote{a} & Date & Sector\tnote{b} & Asymmetry\tnote{c} & WL/NWL\tnote{d} & $v_{\text{asym,max}}$ & $\tau_{\text{asym}}$ & $\tau_{H\alpha}$ & $E_{H\alpha}$ & $E_{\text{bol}}$ & Reference \\
 &  &  &  &  & (km s$^{-1}$)\tnote{e} & (min)\tnote{f} & (min)\tnote{g} & ($10^{30}$ erg)\tnote{h} & ($10^{30}$ erg)\tnote{i} &  \\
\hline
Y1 & 2021 Jan 18 & 34 & symmetry & NWL & - & - & 17 & 0.5 & - & \citetalias{Kajikiya2024} \\
    Y2 & 2021 Jan 19 & 34 & symmetry & NWL & - & - & 21 & 0.8 & - & \citetalias{Kajikiya2024} \\
    Y3 & 2021 Jan 19 & 34 & symmetry & WL & - & - & 37 & 1.3 & 15 & \citetalias{Kajikiya2024} \\
    Y4 & 2021 Jan 24 & 34 & redshift & WL & $402\pm47$ & 319 & 319 & 380 & 12000 & \citetalias{Kajikiya2024} \\
    Y5 & 2021 Jan 25 & 34 & redshift & WL & $265\pm21$ & 35 & 51 & 4.2 & 250 & \citetalias{Kajikiya2024} \\
    Y6 & 2021 Jan 25 & 34 & redshift & WL?\tnote{j} & $295\pm12$ & 6 & 36 & 2.3 & 5.8 & \citetalias{Kajikiya2024} \\
    Y7 & 2021 Jan 25 & 34 & symmetry & WL & - & - & 30 & 1.8 & 40 & \citetalias{Kajikiya2024} \\
    Y8 & 2021 Jan 28 & 34 & blueshift & WL & $445\pm36$ & 8 & 18 & 2.2 & 82 & \citetalias{Kajikiya2024} \\
    Y9 & 2021 Jan 29 & 34 & symmetry & NWL & - & - & 23 & 0.9 & - & \citetalias{Kajikiya2024} \\
    Y10 & 2021 Jan 29 & 34 & symmetry & NWL & - & - & 17 & 0.7 & - & \citetalias{Kajikiya2024} \\
    Y11 & 2021 Jan 30 & 34 & symmetry & NWL & - & - & 14 & 0.6 & - & \citetalias{Kajikiya2024} \\
    Y12 & 2021 Jan 30 & 34 & blueshift & NWL & $254\pm43$ & 14 & 137 & 9.8 & - & \citetalias{Kajikiya2024} \\
    Y13 & 2021 Jan 31 & 34 & symmetry & NWL & - & - & 49 & 1.6 & - & \citetalias{Kajikiya2024} \\
    Y14 & 2021 Jan 31 & 34 & symmetry & NWL & - & - & 8 & 0.2 & - & \citetalias{Kajikiya2024} \\
    Y15 & 2021 Jan 31 & 34 & symmetry & NWL & - & - & 28 & 0.6 & - & \citetalias{Kajikiya2024} \\
    \multirow{2}{*}{Y16\tnote{k}} & \multirow{2}{*}{2021 Jan 31} & \multirow{2}{*}{34} & redshift & \multirow{2}{*}{WL} & $242\pm32$ & 65 & \multirow{2}{*}{146} & \multirow{2}{*}{9.2} & \multirow{2}{*}{19} & \citetalias{Kajikiya2024} \\
     &  &  & redshift &  & $223\pm21$ & 35 &  &  &  & \citetalias{Kajikiya2024} \\
    Y17 & 2021 Feb 2 & 34 & symmetry & NWL & - & - & 26 & 1.5 & - & \citetalias{Kajikiya2024} \\
    Y18 & 2021 Feb 2 & 34 & symmetry & NWL & - & - & 10 & 0.6 & - & \citetalias{Kajikiya2024} \\
    Y19 & 2021 Feb 2 & 34 & symmetry & NWL & - & - & 17 & 0.4 & - & \citetalias{Kajikiya2024} \\
    Y20 & 2021 Feb 2 & 34 & blueshift & NWL & $202\pm18$ & 160 & 192 & 12 & - & \citetalias{Kajikiya2024} \\
    Y21 & 2021 Feb 3 & 34 & redshift & WL & $188\pm27$ & 98 & 122 & 28 & 430 & \citetalias{Kajikiya2024} \\
    Y22 & 2021 Feb 3 & 34 & symmetry & NWL & - & - & 45 & 2.3 & - & \citetalias{Kajikiya2024} \\
    Y23 & 2021 Feb 3 & 34 & symmetry & WL & - & - & 90 & 5.2 & 4.7 & \citetalias{Kajikiya2024} \\
    Y24 & 2021 Feb 3 & 34 & symmetry & NWL & - & - & 30 & 1.2 & - & \citetalias{Kajikiya2024} \\
    Y25 & 2021 Feb 4 & 34 & symmetry & NWL & - & - & 61 & 2.0 & - & \citetalias{Kajikiya2024} \\
    Y26 & 2021 Feb 4 & 34 & symmetry & NWL & - & - & 82 & 3.8 & - & \citetalias{Kajikiya2024} \\
    Y27 & 2021 Feb 4 & 34 & symmetry & WL & - & - & 124 & 7.4 & 66 & \citetalias{Kajikiya2024} \\
\hline
M1 & 2019 Jan 18 & 7 & blueshift & NWL & 95 & 60 & 60 & 4.7 & - & Flare C in \cite{Maehara2021} \\
M2 & 2019 Jan 18 & 7 & symmetry & WL & - & - & 35 & 1.8 & 180 & Flare D in \cite{Maehara2021} \\
\hline
N1 & 2019 Jan 26 & 7 & symmetry & WL & - & - & 90 & 7.8 & 220 & Y1 in \cite{Notsu2024} \\
\multirow{2}{*}{N2\tnote{l}} & \multirow{2}{*}{2019 Jan 27} & \multirow{2}{*}{7} & blueshift & \multirow{2}{*}{NWL} & 200 & 20 & \multirow{2}{*}{258} & \multirow{2}{*}{17} & \multirow{2}{*}{-} & \multirow{2}{*}{Y3 in \cite{Notsu2024}} \\
 &  &  & blueshift &  & 200 & 20 &  &  &  &  \\
N3 & 2019 Jan 28 & 7 & redshift & WL & - & 90 & 66 & 9.3 & 87 & Y4 in \cite{Notsu2024} \\
N4 & 2019 Jan 28 & 7 & symmetry & NWL & - & - & 78 & 3.9 & - & Y5 in \cite{Notsu2024} \\
N5 & 2021 Jan 31 & 34 & symmetry & WL & - & - & 318 & 17 & 36 & Y29 in \cite{Notsu2024} \\
\bottomrule
\end{tabular}
\begin{tablenotes}
\item[a] Label of the H$\alpha$ flare
\item[b] TESS Sector number
\item[c] Classification of asymmetry. ``symmetry" indicates flares with no asymmetry, ``redshift" indicates flares with red asymmetry, and ``blueshift" indicates flares with blue asymmetry.
\item[d] Classification of white-light flares (WL) and non-white-light flares (NWL)
\item[e] Maximum velocity of the asymmetric component (km s$^{-1}$) measured by the fitting process.
\item[f] Duration of asymmetry (min)
\item[g] Duration of the H$\alpha$ flare (min)
\item[h] H$\alpha$ flare energy (erg)
\item[i] Bolometric flare energy (erg)
\item[j] Possible white-light flare with barely detectable white-light enhancement.
\item[k] Y16 shows two red asymmetries during one flare. The first red asymmetry ($\tau_{\text{asym}} \sim 65$ min) occurs simultaneously with the flare peak, while the second ($\tau_{\text{asym}} \sim 35$ min) occurs $\sim$100 minutes after the flare peak.
\item[l] N2 shows two blue asymmetries ($\tau_{\text{asym}} \sim 20$ min $\times$ 2) during one flare. The first blue asymmetry occurs simultaneously with the flare peak, while the second occurs $\sim$60 minutes after the flare peak.
\end{tablenotes}
\end{threeparttable}%
}
\end{table}

\begin{table}[htbp]
    \centering
    \caption{Number of symmetry/asymmetry events in different studies}
    \label{tab:asymmetry_events}
    \begin{tabular}{lccc}
        \hline
        \hline
        Event Category & Redshift & Blueshift & Symmetry \\
        \hline
        Events from \citetalias{Kajikiya2024} & 6 & 3 & 19 \\
        Events from \citet{Maehara2021} & 0 & 1 & 1 \\
        Events from \citet{Notsu2024} & 1 & 2 & 3 \\
        All Events & 7 & 6 & 23 \\
        Events with $E_{\mathrm{H\alpha}} > 2 \times 10^{30}$ erg & 7 & 6 & 8 \\
        \hline
    \end{tabular}
\end{table}


\section{Asymmetry vs Flare properties} \label{sec:Asymmetries vs Flare properties}
In this section, we discuss the dependence of the H$\alpha$ line profile asymmetry on energy, duration of H$\alpha$ flare, and its association with white-light flares.

\subsection{Dependence of H$\alpha$ Asymmetry on Energy and Duration}\label{sec:Dependence of Hα Asymmetry on Energy and Duration}
The flare energy and duration reflect the scale of the flare and its impulsiveness. In the case of the Sun, it is known that impulsiveness (high heating rate per area) induces strong white-light emissions \citep{Watanabe2017} and/or high chromospheric condensation velocity \citep{Longcope2014,Tian2015}. Also, \cite{Watanabe2017} mentioned that impulsiveness can be related to the coronal magnetic field strength of the flaring region, which may be related to the occurrence of prominence eruptions. Therefore, we may expect that there could be differences in the distribution of blue/red asymmetry and white-light/non-white-light events in the diagram of flare energy and duration.

\begin{figure}
    \centering
    \includegraphics[width=1\linewidth]{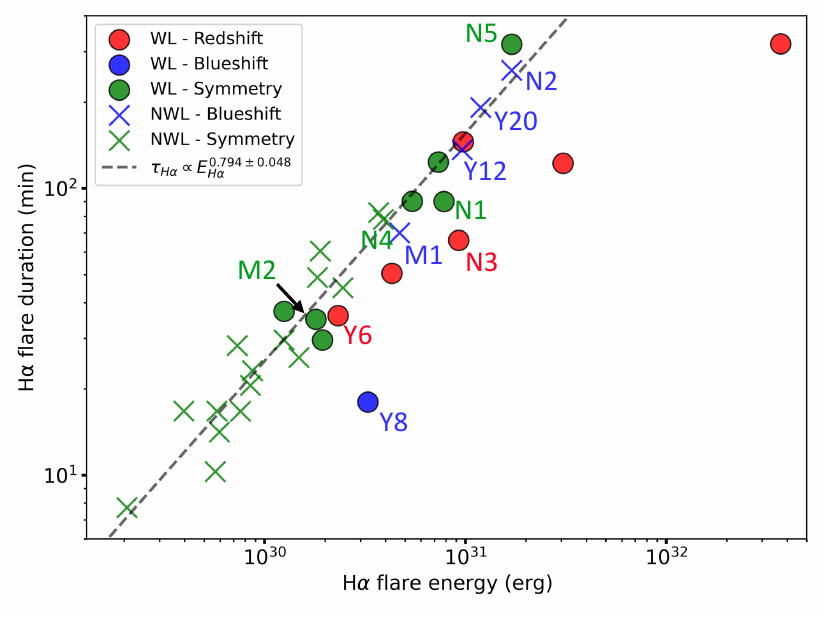}
    \caption{Relation between H$\alpha$ flare energies and durations. The difference in color represents the asymmetry type, while the difference in marker style represents whether the flares are white-light flares or non-white-light flares. The black dashed line indicates the fitted line for only symmetry events.
    }
    \label{f:Energy_Duration}
\end{figure}

Figure \ref{f:Energy_Duration} shows the relation between H$\alpha$ flare energy and duration. This shows that flares showing significant asymmetry have energies greater than 2 $\times$ 10$^{30}$ erg. This energy may represent the detection limit for asymmetry in the sample used in this study. For low-energy flares, the flare-related emission is relatively weak, making it difficult to distinguish asymmetric components. Above this energy threshold, both red and blue asymmetry events show a wide range of energies and durations. Figure \ref{f:Energy_Duration} shows that, above this threshold of 2 $\times$ 10$^{30}$ erg, flares with higher energies tend to show asymmetry more frequently. Among flares with energies above 9 $\times$ 10$^{30}$ erg, 7 out of 8 show asymmetry, while in the energy range of 2–9 $\times$ 10$^{30}$ erg, 4 out of 10 show asymmetry.
Also, some red asymmetry events seem to show slight deviations from the fitted line for symmetry events and have relatively shorter durations compared to symmetry events. If this is true, this property is consistent with solar chromospheric condensation, where a higher energy deposition rate induces faster chromospheric condensation (e.g., \citealt{Longcope2014}). However, the trend is not necessarily evident, and the sample number is not statistically enough. We then need to be careful with this interpretation.
Finally, Figure \ref{f:Energy_Duration} also suggests that there is no clear difference of flare impulsiveness between blue asymmetry events and symmetry events. It is, however, interesting that the Y8 showing blue asymmetry has exceptionally shorter durations compared to other flares.



\subsection{Dependence of H$\alpha$ Asymmetry on Association with White-Light Flares}
White-light flares are thought to be represent radiation mainly from the flare ribbons, which can originate from chromospheric condensation during the impulsive phase. We consider that the existence of white-light emissions is a key to understand the origin of H$\alpha$ asymmetry of flares in the following two reasons: White-light emissions account for the majority of the total energy of the flare. \citep{Kretzschmar2010,Watanabe2013,Osten2015,Namekata2017}.
Second, white-light radiation is strongly correlated with non-thermal radiation, such as hard X-rays \citep{Watanabe2010} and is a good indicator of the timescale and magnitude of energy release. Therefore, comparing the asymmetry type in the H$\alpha$ line with the white-light enhancement is useful for understanding the origins of these asymmetry.
In the case of the Sun, the difference between white-light flare and non-white-light flare depends on impulsiveness of flares. White-light flares show higher rates of non-thermal energy deposition compared to non-white-light flares \citep{Watanabe2017}. Therefore, flares with small heating rates may not be detected in the white light and could be classified as non-white-light flares in the stellar case. Additionally, since white-light flares originate from the flare footpoints, white-light flares that occur near the limb may not be detected and could be classified as non-white-light flares. 

Before going into the discussion of the relationship with asymmetry, we mention the detection threshold of white-light flares. Figure \ref{f:Energy_Duration} show white-light flares have H$\alpha$ energies greater than 10$^{30}$ erg. In TESS Cycle 1 and Cycle 3 observations of YZ CMi, flares with bolometric energies exceeding 10$^{32}$ erg follow the energy-frequency distribution and can be reliably detected \citep{Maehara2021,Ikuta2023}. The H$\alpha$ flare energies are typically $\sim$0.01 of the bolometric energy reported in \citetalias{Kajikiya2024}. Therefore, white-light emission for H$\alpha$ flare energies below  $\sim$10$^{30}$ erg would be too weak to be detected by TESS. 

\begin{figure}
    \centering
    \includegraphics[width=1\linewidth]{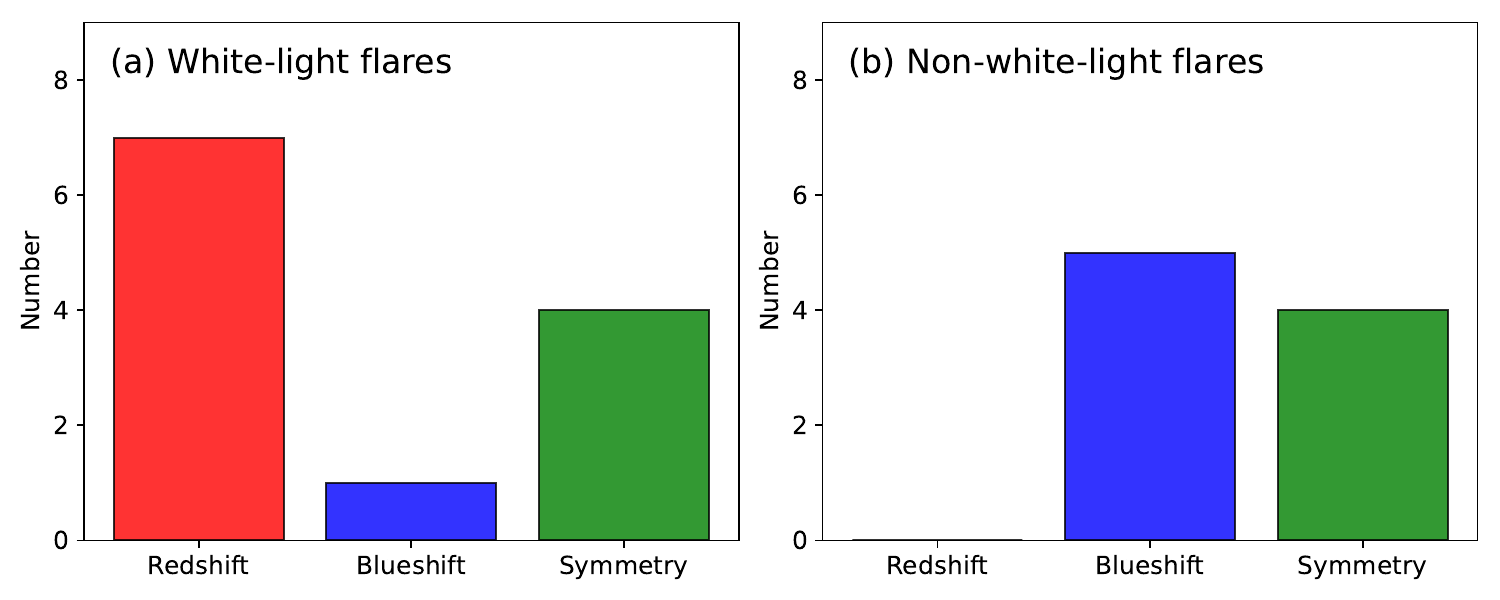}
    \caption{Comparison of histograms for (a) white-light flares and (b) non-white-light flares categorized by asymmetry type. The vertical axis represents the number of asymmetry events (Events with $E_{\mathrm{H\alpha}} > 2 \times 10^{30}$ erg in Table \ref{tab:asymmetry_events}). We note that if one flare exhibits multiple independent asymmetries, each asymmetry is counted separately.}
    
    \label{f:WL_dependence}
\end{figure}

Figure \ref{f:WL_dependence} (a) and (b) show the number of detected asymmetry events for white-light flares and non-white-light flares, respectively. Only events with H$\alpha$ flare energies greater than 2 $\times$ 10$^{30}$ erg (Events with $E_{\mathrm{H\alpha}} > 2 \times 10^{30}$ erg in Table \ref{tab:asymmetry_events}) are displayed, as this energy may represent the detection limit for line asymmetry (see Figure \ref{f:Energy_Duration}).
Figure \ref{f:WL_dependence} shows that all 7 events showing red asymmetry are associated with white-light flares. Additionally, this trend is statistically significant based on Fisher's exact test (p-value = 0.007), as described in Appendix \ref{Appendix:Fisher's_Exact_Test}. This result suggests that the origins of red asymmetry are related to white-light flares. 
This relation between white-light flares and red asymmetry is consistent with the properties of chromospheric condensations, as chromospheric condensations are one origin of white-light flares \citep{Kowalski2024}. On the other hand, these relations are not consistent with a backward prominence eruption (e.g., \citetalias{Kajikiya2024}). Backward prominence eruptions occur on the stellar limb, and thus, a white-light flares associated with such an eruption may be less observable and could be classified as a non-white-light flares. These support the hypothesis that red asymmetry originates from the chromospheric condensation in most cases.

For the blue asymmetry, Figure \ref{f:WL_dependence} (a) and (b) show 5 out of 6 blue asymmetry events are non-white-light flares. This trend is statistically significant based on Fisher's exact test (p-value = 0.029) similar to red asymmetry, as described in Appendix \ref{Appendix:Fisher's_Exact_Test}. The energies of flares showing blue asymmetry are higher than the TESS detection threshold of $E_{H\alpha}>2\times 10^{30}$ erg. Additionally, The previous paper discussed that these blue asymmetries might be caused by prominence eruptions (\citealt{Maehara2021,Notsu2024}; \citetalias{Kajikiya2024}). \footnote{We note that one red asymmetry event (Y6 in \citetalias{Kajikiya2024}) suggesting prominence eruption shows a weak white-light enhancement that barely clears the flare detection criteria in \citetalias{Kajikiya2024}.} 

There are two possible reasons for a flare being classified as a non-white-light flare. First, the flare has a low heating rate and does not emit detectable white-light radiation, similar to non-white-light flares observed on the Sun.
\cite{Watanabe2017} suggest that white-light flares occur in flares with $\sim$10–30 G stronger coronal magnetic fields at the energy-release site compared to non-white-light flares. If these differences also apply to M-dwarfs, the stronger magnetic fields for white-light flares may suppress prominence eruptions more strongly \citep{Drake2016,Alvarado-Gomez2018,Sun2022}. This might lead to a relatively higher association rate of prominence eruption for non-white-light flares.

\cite{Kazachenko2023} reported that eruptive solar flares tend to have longer durations than confined flares in GOES X-ray data. Additionally, non-white-light flares show longer durations than white-light flares in GOES X-ray data \citep{Watanabe2017}. 
These solar observations suggest non-white-light flares may show higher association rate of prominence eruption compared to white-light flares. Therefore, the higher association rate of prominence eruption candidates for non-white-light flares on M-dwarfs might not be surprising based on solar properties. On the other hand, as discussed in Section \ref{sec:Dependence of Hα Asymmetry on Energy and Duration}, no clear difference in flare impulsiveness between blue asymmetry events and symmetry events is observed. Therefore, the hypothesis that flares with a low heating rate have a higher association rate with prominence eruptions is not significantly validated based on the H$\alpha$ flare energy and duration. We note that in the discussion here, we assume that the flare duration trends observed in GOES X-ray data can be applied to the H$\alpha$ line based on its similarity, but the potential difference between the two should be investigated in future studies.

The second reason why flares might not be classified as white-light flares is that they occurred on the stellar limb. White-light emission may not be observable when the flare footpoints are obscured near the limb. 
In the case of the Sun, white-light flares tend to show 
reduced white-light flux near the limb, even when the footpoints are not completely obscured \citep{Kuhar2016}, although the mechanism is not well understood. While it remains unclear whether this limb-darkening-like effect applies to M-dwarf flares, white-light flares occurring near the limb might be less observable on M-dwarfs, even if the footpoints are not completely obscured. As a result, they may potentially be classified as non-white-light flares. As discussed in \citetalias{Kajikiya2024}, prominence eruptions on the disk would be less observable due to their lower contrast with the background compared to those on the limb on M-dwarfs \citep{Leitzinger2022}. This may have reduced the detection rate of prominence eruptions on the disk compared to those on the limb. This potential observational bias could explain the relatively higher association rate of prominence eruptions for non-white-light flares, assuming that most prominence eruption candidates have occurred on the limb. Additionally, this bias may also explain the lower association rate of prominence eruptions on M-dwarfs ($\sim$20$\%$; \citealt{Notsu2024}; \citetalias{Kajikiya2024}) compared to an active G-dwarf ($\sim$60$\%$; \citealt{Namekata2024a}), as on-disk eruptions on G-dwarfs can be detected as blue-shifted absorption \citep{Namekata2022}. Moreover, the observed line-of-sight velocities can be slower on the limb than on the disk due to the projection effect. For example, a prominence eruption ejected at an angle of $\sim$70 degrees from the line of sight has a true velocity that is $\sim$2 times the radial velocity ($\sim$200--300 km s$^{-1}$). These true velocities ($\sim$400--600 km s$^{-1}$) can be comparable to the eruption velocities observed on G-dwarfs \citep{Namekata2022,Namekata2024a}. This potentially fill the gap in the scaling relations for flare energy and kinetic energy between the Sun and M-dwarfs (e.g., \citealt{Moschou2019}). 
In the Section \ref{sec:Relation Between Starspots and Prominence Eruptions}, we examine whether the starspots that could be the origin of these prominence eruptions are located on the limb.

\section{Asymmetry vs Spot Properties
}\label{sec:Asymmetries vs Spot Properties}
\subsection{Dependence of White-Light Flare Occurrence Frequency on Rotational Phase}\label{subsec:WLF Dependence on Rotational Phase}
\begin{figure}
    \centering
    \includegraphics[width=0.5\linewidth]{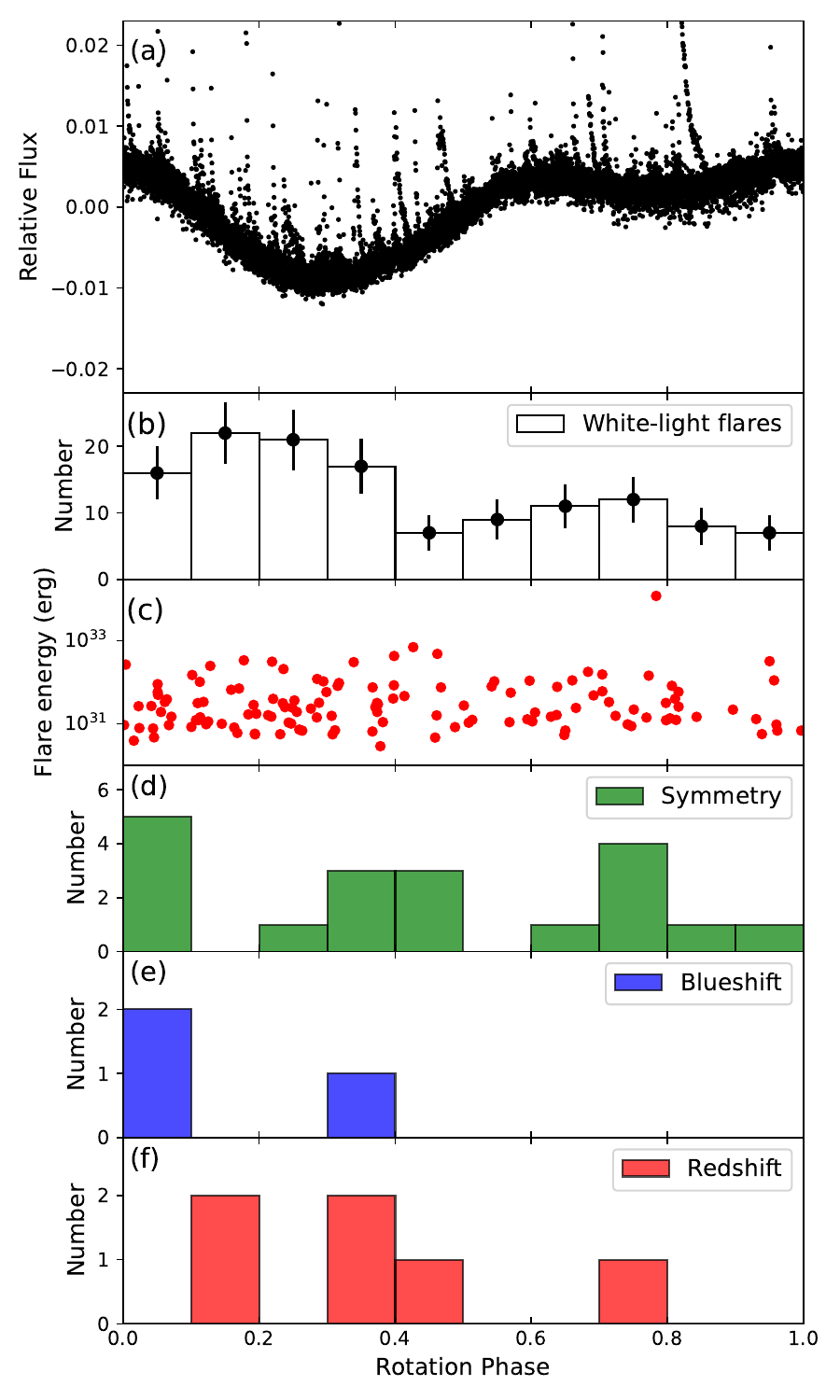}
    \caption{The panel (a) shows the light curve folded with the period (2.774 days) from BJD 2458402.09. The panels (b) and (c) show histograms for detected white-light flares and flare bolometric energy of each event as a function of rotational phase. The panels (d), (e), and (f) show histograms of detected symmetry and asymmetry events (Events from \citetalias{Kajikiya2024} in Table \ref{tab:asymmetry_events}). If one flare exhibits multiple independent asymmetries, each asymmetry is counted separately, similar to Figure 1.}
    \label{f:Phase_dependence}
\end{figure}

Some previous statistical researches have shown that the occurrence frequency of white-light flares does not depend on the rotational phase on M-dwarfs (e.g., \citealt{Hawley2014,Maehara2021,Ikuta2023}). \cite{Bicz2022} reported a dependence of white-light flare frequency on the rotational phase of YZ CMi using the 2-minute cadence data from TESS Cycle 3. On the other hand, \cite{Ikuta2023}, using the same data as \cite{Bicz2022}, cannot conclude such dependence\footnote{We found that the reduced chi-square value and p-value in \cite{Ikuta2023} are incorrect due to an error of the calculation. We correctly calculate these values to be 0.906 (0.464) and 0.519 (0.899) for all flares (for flares with longer equivalent duration than 1s).
The corrected p-value is still larger than the significance level of 0.05, so the conclusion in \cite{Ikuta2023} remains unchanged.
}. As seen in these, the dependence of white-light flare frequency on the rotational phase remains ambiguous. In \citetalias{Kajikiya2024}, we conducted flare detection using the 20-second cadence data from TESS Cycle 3, unlike \cite{Bicz2022} and \cite{Ikuta2023}. Figure \ref{f:Phase_dependence} (a), (b), and (c) show the TESS light curve, a histogram of detected flares as a function of the rotational phase, and the bolometric energy of detected white-light flares, respectively. Figure \ref{f:Phase_dependence} (b) shows a tendency for a higher frequency of white-light flares phase at the phases 0.1--0.4, when large starspots are visible on the disk center \citep{Ikuta2023}. 
We assume a null hypothesis that the flares are uniformly distributed across 10 bins of the rotation phase and performed a chi-squared test with 9 degrees of freedom \citep{Pearson1900}, following the same method as \cite{Ikuta2023}, to assess whether this trend is statistically significant. We calculated the chi-square $\chi^2$ using the following equation:
\begin{equation}\label{eq:kai_square_test}
\chi^2 = \sum_{i=1}^{10} \frac{(O_i - E_i)^2}{E_i}
\end{equation}
where $O_i$ is the number of detected flares in each bin. $E_i$ represents the expected number of flares in each bin under a uniformly distributed flares per time:  
\begin{equation}
E_i = \frac{N_{\text{total}} \times T_i}{T_{\text{total}}}
\end{equation}
where $N_{\text{total}}$ is the total number of observed flares (130 events), $T_i$ is the observation time in each bin, and $T_{\text{total}}$ is the total observation time (23.456 days). As a result of the chi-squared test, the $\chi^2$ value and p-value are calculated to be 17.70 and 0.039. This p-value is less than the significance level of 0.05, indicating that the null hypothesis is rejected with statistical significance. This suggests that the frequency of flares correlates with the rotational phases corresponding to the projected area of starspots. At phases of 0.6--1.0, the light curve also shows a slight decrease in brightness ($\sim$0.05\%) caused by starspots. The difference in flare occurrence frequency depending on the rotational phase suggests that larger spots produce more frequent flares.

It should be noted that our chi-squared result is consistent with \cite{Bicz2022} but differs from \cite{Ikuta2023}. 
This difference is due to the number of detected flares in each bin.
Unlike this work and \cite{Bicz2022}, \cite{Ikuta2023} used the machine learning method by \cite{Feinstein2020} to detect flares for data with two-minute cadence. This machine learning method may have a detection sensitivity dependence on the rotational phase because it uses the light curve for flare detection without detrending stellar rotational variations, although flares detected with detrending method are adopted as the train data of the machine learning.
\subsection{Dependence of H$\alpha$ Line Asymmetry on Rotational Phase}
We investigated the dependency of the occurrence frequency of blue/red asymmetries on the rotational phase using the data only in \citetalias{Kajikiya2024} (Events from \citetalias{Kajikiya2024} in Table \ref{tab:asymmetry_events}). Figure \ref{f:Phase_dependence} (d), (e), and (f) show the histograms of the number of events showing no asymmetry, red asymmetry, and blue asymmetry, respectively. These figures show 8 out of 9 blue/red asymmetries occur during phases 0--0.5, while only 1 of 9 occurs during phases 0.5--1.0. The ratio of flares showing asymmetry to the total H$\alpha$ flares is 7 out of 19 during phases 0--0.5 and 1 out of 8 during phases 0.5--1.0. These results suggest that the occurrence of these asymmetries depends on the rotational phase. On the other hand, this trend is not statistically significant based on the result of Fisher's exact test (p-value = 0.078), as described in Appendix \ref{Appendix:Fisher's_Exact_Test}.

Figure \ref{f:Phase_dependence} shows that 5 out of 6 red asymmetry events occur at phase 0.1–0.5. The rate of H$\alpha$ flares showing red asymmetry is 4 out of 19 at the phases 0.1–0.5, whereas it is only 1 out of 8 at all the other phases. Although the sample size is small, this higher occurrence of red asymmetry at the phases 0.1–0.5 suggests that the origin of red asymmetry is related to large spots, similar to white-light flares. In the case of the Sun, chromospheric condensation is more observable on the disk compared to the limb, as these originate from flare footpoints \citep{Svestka1962}. Therefore, assuming these red asymmetry events originate from large starspots, their higher occurrence rate of H$\alpha$ flares showing red asymmetries at the phases 0.1--0.5 is consistent with chromospheric condensation. However, the probability of an asymmetry event occurring does not show a statistically significant difference between phase 0–0.5 and phase 0.5–1 based on Fisher's exact test (see Appendix \ref{Appendix:Fisher's_Exact_Test}). We then need to be cautious in interpreting this result.

\subsection{Relation Between Starspots and Prominence Eruption Candidates}\label{sec:Relation Between Starspots and Prominence Eruptions}
In the previous section, we investigated the statistical characteristics of prominence eruption candidates in relation to flare properties. As a result, we found that prominence eruption candidates show a low association rate with white-light enhancement. 
As one possibility, we suggest that this result may mean that most prominence eruption candidates originate on the limb, as discussed in Section \ref{sec:Asymmetries vs Flare properties}. 
To examine the statistical dependencies of prominence eruption candidates on their potential origin in starspots, we derive the spot properties at the time of all 7 prominence eruption candidates (Table \ref{tb:prom_yzcmi_c3_three_spot} and \ref{tb:prom_yzcmi_c1_two_spot}) using the result of starspot mapping for the TESS Cycle 1 and 3 light curves estimated by \citet{Ikuta2020,Ikuta2023}. The detailed model description and spot parameters are provided in Appendix \ref{Appendix:Spot Model Description}. The properties of all prominence eruption candidates are listed in Table \ref{tab:Flares_spot_discussion}. 

\begin{figure}
    \centering
    \includegraphics[width=1\linewidth]{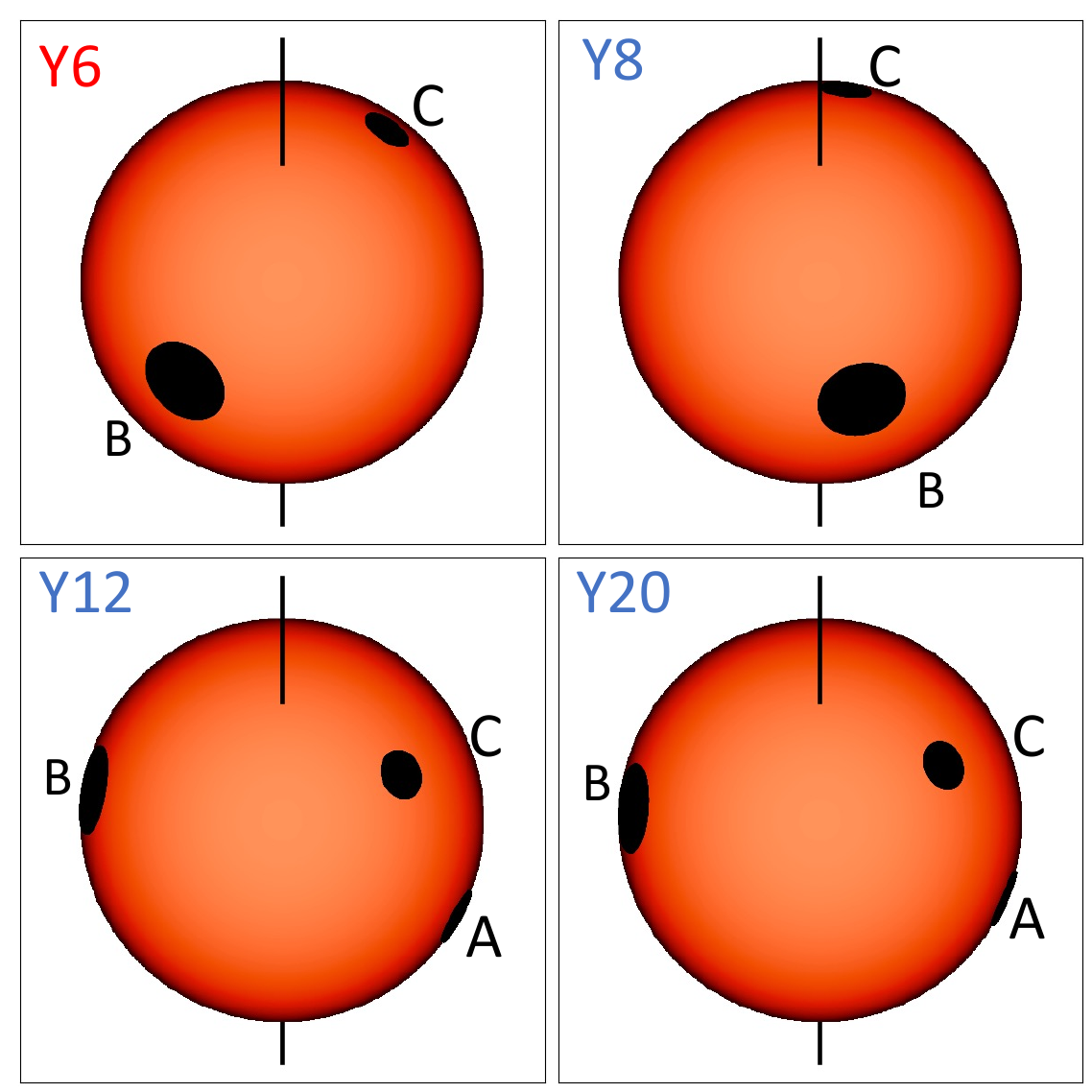}
    \caption{Starspot map at the time of prominence eruption candidates from TESS Cycle 3 light-curve modeling.
    }
    \label{f:spot_model_c3}
\end{figure}

\begin{figure}
    \centering
    \includegraphics[width=1\linewidth]{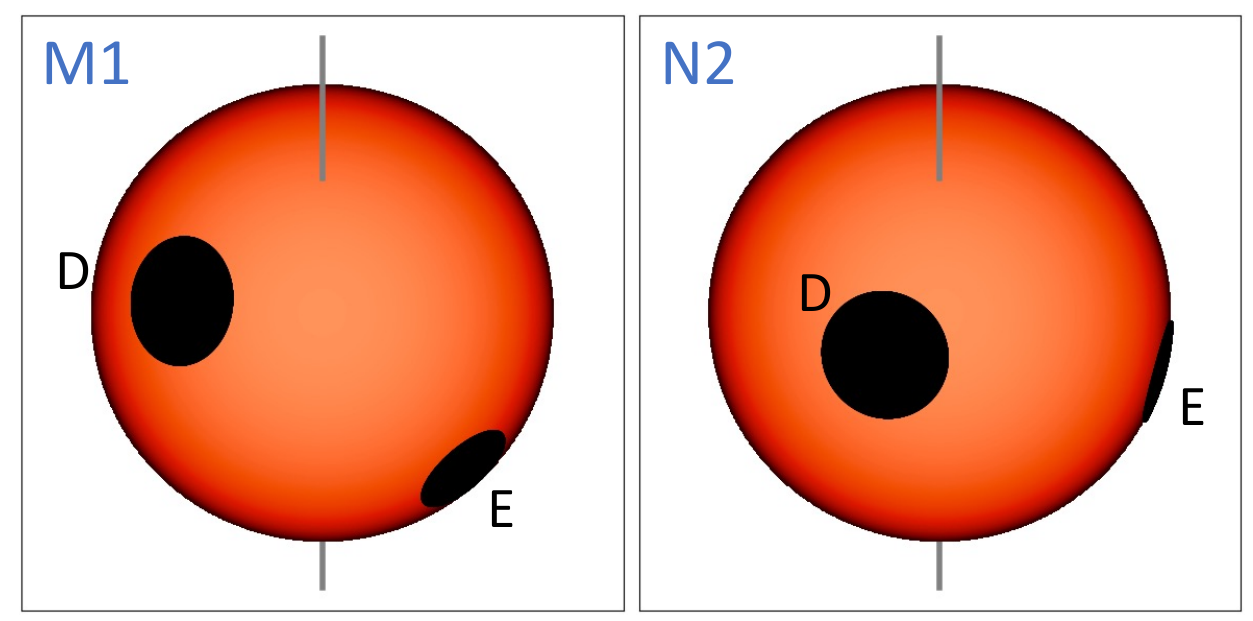}
    \caption{The same as Figure 4, but for TESS Cycle 1. We note that N2 shows two independent blue asymmetries, suggesting prominence eruptions.
    }
    \label{f:spot_model_c1}
\end{figure}

Figures \ref{f:spot_model_c3} and \ref{f:spot_model_c1} show the starspot distributions at the times of prominence eruption candidates in TESS Cycle 3 and Cycle 1, respectively. These figures suggest that starspots are located on the limb for each event, although the spot mapping can reflect only the global distribution of starspots and might not reflect all small spots and polar spots (cf. Appendix \ref{Appendix:Spot Model Description}). Assuming that the flares and prominence eruptions originated from the spots estimated by spot mapping, this result is consistent with the hypothesis that most prominence eruption candidates occurred on the limb.
On the other hand, starspots are also located on the disk center for each events. Therefore, we cannot distinguish whether these events occurred on the limb or disk center based on starspot distributions. 
Further starspot mapping at the time of prominence eruptions and simultaneous Zeeman Doppler Imaging \citep{Namekata2024b}, is necessary to investigate the potential role of magnetic suppression and differences in the visibility of prominence eruptions between on the disk and on the limb.
\section{Summary and Conclusions}\label{sec:Summary and Conclusions}
In \citetalias{Kajikiya2024}, we reported 27 H$\alpha$ flares observed on the active M-dwarf YZ CMi during TESS Cycle 3, among which 5 show red asymmetries and 3 show blue asymmetries in the H$\alpha$ line profile. 
We discovered 4 prominence eruption candidates, of which 3 show blue asymmetries and 1 shows a red asymmetry. In this study (Paper II), we investigated the statistical characteristics of H$\alpha$ line asymmetries in relation to flare and spot properties using these data, as well as previously reported simultaneous observation data of 7 events (cf. Table \ref{tab:Flares_spot_discussion}) with TESS Cycles 1 and 3.

As a result, we found that flares with higher energies more frequently show blue/red asymmetries. Additionally, we found the frequency of asymmetry events depends on the association with white-light flares. For red asymmetries, all 7 red asymmetry events are associated with white-light flares, consistent with chromospheric condensation. On the other hand, for blue asymmetries, 5 out of 6 events are not associated with white-light flares. We note that all these blue asymmetry events and one red asymmetry event, which show only barely detectable white-light enhancement, are candidates for prominence eruptions. 

Next, for spot properties, we found that the frequencies of white-light flares and the association rate of red asymmetries depend on the stellar rotational phase. These events occur more frequently during phases suggesting the presence of large starspots on the disk. For blue asymmetries, although the sample size is small, 2 out of 3 events in Paper I occurred during the rotational phases suggesting large starspots located on the limb. We further explored the starspots distribution at the time of prominence eruption candidates using spot model. We found that all prominence eruption candidates might have occurred when starspots were located not only on the disk but also on the limb.

Based on these characteristics, we propose two possible hypotheses. The first hypothesis is that flares with lower heating rates have a higher association rate with prominence eruptions. The second is that most detected prominence eruptions occurred on the limb. The latter hypothesis may explain the low association rate of prominence eruptions with H$\alpha$ flares on M-dwarfs (e.g., \citealt{Kajikiya2024}) and fill the gap in the scaling relations for flare energy and kinetic energy between the Sun and M-dwarfs (e.g., \citealt{Moschou2019}), as discussed in this paper.

\section*{Acknowledgment}
We thank Kazunari Shibata for his valuable comments and discussions.
This research was supported by JSPS (Japan Society for the Promotion of Science) KAKENHI Grant Numbers  JP21J00316 (K.N.), JP20K04032, JP24K00685 (H.M.), JP24K17082 (K.I.), JP24K00680 (K.N., H.M., and D.N.), and JP24H00248 (Y.K., K.N., K.I., H.M., and D.N). Y.N. acknowledges the funding support from NASA ADAP 80NSSC21K0632, and NASA TESS Cycle 6 80NSSC24K0493.
The spectroscopic data used in this paper were obtained through the program 21A-N-CN03 (PI: K.N.) with the 3.8m Seimei telescope, which is located at Okayama Observatory of Kyoto University.
This paper includes data collected with the TESS mission, obtained from the MAST data archive at the Space Telescope Science Institute (STScI). Funding for the TESS mission is provided by the NASA Explorer Program.
STScI is operated by the Association of Universities for Research in Astronomy, Inc., under NASA contract NAS 5-26555. 
Some of the data presented in this paper were obtained from the Mikulski Archive for Space Telescopes (MAST) at the Space Telescope Science Institute. The specific observations analyzed can be accessed via\dataset[10.17909/14ym-zt14]{https://doi.org/10.17909/14ym-zt14}. The authors acknowledge ideas from the participants in the workshop ``Blazing Paths to Observing Stellar and Exoplanet Particle Environments" organized by the W.M. Keck Institute for Space Studies.

\facilities{Seimei telescope, Transiting Exoplanet Survey Satellite (TESS)}

\software{\textsf{astropy} \citep{Astropy2018} , \textsf{IRAF} \citep{IRAF1986}, \textsf{PyRAF} \citep{Pyraf2012}}

\clearpage

\appendix
\section{Appendix: Fisher's Exact Test}\label{Appendix:Fisher's_Exact_Test}
We performed Fisher's exact test \citep{Fisher1922} to assess the dependence of blue/red asymmetry on white-light flares and rotational phase. Since the sample size of H$\alpha$ flares is small, the chi-squared test does not have sufficient statistical power, unlike for white-light flares. Therefore, we used Fisher's exact test for H$\alpha$ flares.

To examine the dependence of blue/red asymmetry on white-light flares, as shown in Figure \ref{f:WL_dependence}, we performed two Fisher's exact tests. For red asymmetry, we assumed the null hypothesis that the probability of red asymmetry event occurring is the same for white-light flares and non-white-light flares. We set the alternative hypothesis that red asymmetry event is more likely to occur in white-light flares than in non-white-light flares. For blue asymmetry, we assumed the null hypothesis that the probability of blue asymmetry event occurring is the same for white-light flares and non-white-light flares. We set the alternative hypothesis that blue asymmetry event is more likely to occur in non-white-light flares than in white-light flares. Tables \ref{tab:red_asymmetry_wl_nwl} and \ref{tab:blue_asymmetry_wl_nwl} present the contingency tables for Fisher's exact tests of red asymmetry and blue asymmetry events, respectively. As a result of the Fisher's exact tests, the p-values are 0.007 and 0.029, respectively. These values are below 0.05, leading to the rejection of both null hypotheses. This result indicates that red asymmetry events are more likely to occur in white-light flares than in non-white-light flares, while blue asymmetry events are more likely to occur in non-white-light flares than in white-light flares.

Next, we performed three Fisher's exact tests to assess the dependence of blue/red asymmetry on rotational phase, as shown in Figure \ref{f:Phase_dependence}. We binned the phases into 0–0.5 and 0.5–1, as listed in Tables \ref{tab:asymmetry_phase}--\ref{tab:blue_asymmetry_phase}. In the first test, we assumed the null hypothesis that the probability of an asymmetry event occurring is the same for phase 0–0.5 and phase 0.5–1. We set the alternative hypothesis that asymmetry events are more likely to occur in the phase 0–0.5 than in the phase 0.5–1. In the second and third tests, we applied the same Fisher's exact tests separately to red asymmetry events and blue asymmetry events. Tables \ref{tab:asymmetry_phase}--\ref{tab:blue_asymmetry_phase} present these contingency tables. As a result of the Fisher's exact tests, all p-values are above 0.05, meaning that the null hypotheses were not rejected, as listed in Tables \ref{tab:asymmetry_phase}--\ref{tab:blue_asymmetry_phase}.


\begin{table}[h]
    \centering
    \caption{Red Asymmetry Dependence on WL/NWL ($E_{H\alpha} > 2 \times 10^{30}$ erg)}
    \label{tab:red_asymmetry_wl_nwl}
    \begin{tabular}{ccc}
        \hline
        \hline
        & WL & NWL \\
        \hline
        Red Asymmetry & 7 & 0 \\
        Others & 5 & 9 \\
        \hline
        p-value & & \textbf{0.007} \\
        \hline
    \end{tabular}
\end{table}

\begin{table}[h]
    \centering
    \caption{Blue Asymmetry Dependence on WL/NWL ($E_{H\alpha} > 2 \times 10^{30}$ erg)}
    \label{tab:blue_asymmetry_wl_nwl}
    \begin{tabular}{ccc}
        \hline
        \hline
        & WL & NWL \\
        \hline
        Blue Asymmetry & 1 & 5 \\
        Others & 11 & 4 \\
        \hline
        p-value & & \textbf{0.029} \\
        \hline
    \end{tabular}
\end{table}

\begin{table}[h]
    \centering
    \caption{Asymmetry Dependence on Rotational Phase}
    \label{tab:asymmetry_phase}
    \begin{tabular}{ccc}
        \hline
        \hline
        & Phase 0.0--0.5 & Phase 0.5--1.0 \\
        \hline
        Red/Blue Asymmetry & 8 & 1 \\
        Others & 9 & 8 \\
        \hline
        p-value & & 0.078 \\
        \hline
    \end{tabular}
\end{table}

\begin{table}[h]
    \centering
    \caption{Red Asymmetry Dependence on Rotational Phase}
    \label{tab:red_asymmetry_phase}
    \begin{tabular}{ccc}
        \hline
        \hline
        & Phase 0.0--0.5 & Phase 0.5--1.0 \\
        \hline
        Red Asymmetry & 5 & 1 \\
        Others & 12 & 8 \\
        \hline
        p-value & & 0.296 \\
        \hline
    \end{tabular}
\end{table}

\begin{table}[h]
    \centering
    \caption{Blue Asymmetry Dependence on Rotational Phase}
    \label{tab:blue_asymmetry_phase}
    \begin{tabular}{ccc}
        \hline
        \hline
        & Phase 0.0--0.5 & Phase 0.5--1.0 \\
        \hline
        Blue Asymmetry & 3 & 0 \\
        Others & 14 & 9 \\
        \hline
        p-value & & 0.262 \\
        \hline
    \end{tabular}
\end{table}

\section{Appendix: Description of the Starspot Model}\label{Appendix:Spot Model Description}

We briefly describe the method and result of the starspot mapping for YZ CMi in Section \ref{sec:Asymmetries vs Spot Properties}. In \cite{Ikuta2023}, we optimize the TESS light curves of YZ CMi in its Sector 7 (Cycle 1) and 34 (Cycle 3) with the analytical model of the spotted flux \citep{Kipping2012} implemented in the code \citep{Ikuta2020} and compare the number of spots by calculating of the model evidence. The modeled flux is specified in the stellar equatorial period, degree of differential rotation, spot latitude, longitude, and radius, under a certain number of spots.
We respectively adopt one/two- and two/three-spot models for the TESS light curves in Cycle 1 and 3 because the light curves show one and two local minima per rotational period. 
As a result, the parameters are deduced for each of the models, and two- and three-spot models are preferred because more number of spots increases the model evidence.
In this paper, we discuss the spot properties as the origin of the asymmetry associated with stellar flares, and as in \cite{Namekata2024b} we derive the spot locations and sizes on the time of the flares (Table \ref{tb:prom_yzcmi_c1_two_spot} and \ref{tb:prom_yzcmi_c3_three_spot}).

We note that \cite{Bicz2022} also derives spot properties from the TESS light curves and results in different spot map from \cite{Ikuta2023}. This is because the result depends on the assumptions of the spot mapping, and different number of spots can reproduce almost similar light curve \citep{Ikuta2020}.


\begin{deluxetable*}{lcccc}
\tablecaption{Spot parameters of YZ CMi Cycle 3 for the Event Y6, Y8, Y12, and Y20 in Paper I} \label{tb:prom_yzcmi_c3_three_spot}
\tabletypesize{\scriptsize}
\tablehead{
\colhead{Parameters} & \colhead{Y6} & \colhead{Y8} & \colhead{Y12} & \colhead{Y20}
}
\startdata
(Spot A) & & & & \\
Latitude (deg) & \multicolumn{4}{c}{$ -10.92 ^{+ 0.50 }_{- 0.45 }$} \\
Longitude (deg) & $ -95.57 ^{+ 0.06 }_{- 0.11 }$ & $ 62.45 ^{+ 0.74 }_{- 0.69 }$ & $ -88.75 ^{+ 0.11 }_{- 0.11 }$ & $ 69.50 ^{+ 0.69 }_{- 0.74 }$ \\
Radius (deg) & \multicolumn{4}{c}{$ 7.91 ^{+ 0.14 }_{- 0.19 }$} \\
Radius ($R_{\rm star}$) & \multicolumn{4}{c}{$ 0.138 ^{+ 0.002 }_{- 0.003 }$} \\
Radius ($10^{10}$ cm) & \multicolumn{4}{c}{$ 0.36 ^{+ 0.02 }_{- 0.07 }$} \\ \hline
(Spot B) & & & & \\
Latitude (deg) & \multicolumn{4}{c}{$ 16.36 ^{+ 0.36 }_{- 0.29 }$} \\
Longitude (deg) & $ 93.33 ^{+ 0.29 }_{- 0.29 }$ & $ -89.32 ^{+ 0.34 }_{- 0.57 }$ & $ 99.29 ^{+ 0.34 }_{- 0.29 }$ & $ -82.51 ^{+ 0.40 }_{- 0.52 }$ \\
Radius (deg) & \multicolumn{4}{c}{$ 12.34 ^{+ 0.05 }_{- 0.05 }$} \\
Radius ($R_{\rm star}$) & \multicolumn{4}{c}{$ 0.215 ^{+ 0.001 }_{- 0.001 }$} \\
Radius ($10^{10}$ cm) & \multicolumn{4}{c}{$ 0.56 ^{+ 0.04 }_{- 0.10 }$} \\ \hline
(Spot C) & & & & \\
Latitude (deg) & \multicolumn{4}{c}{$ 49.48 ^{+ 5.76 }_{- 2.69 }$} \\
Longitude (deg) & $ -166.90 ^{+ 2.46 }_{- 2.46 }$ & $ -125.13 ^{+ 2.35 }_{- 2.52 }$ & $ 132.12 ^{+ 2.12 }_{- 2.75 }$ & $ 136.94 ^{+ 2.46 }_{- 2.41 }$ \\
Radius (deg) & \multicolumn{4}{c}{$ 6.59 ^{+ 0.41 }_{- 0.18 }$} \\
Radius ($R_{\rm star}$) & \multicolumn{4}{c}{$ 0.115 ^{+ 0.007 }_{- 0.003 }$} \\
Radius ($10^{10}$ cm) & \multicolumn{4}{c}{$ 0.30 ^{+ 0.02 }_{- 0.05 }$} \\
\enddata
\end{deluxetable*}

\begin{deluxetable*}{lcc}
\tablecaption{Spot parameters of YZ CMi Cycle 1 for the M1 in \cite{Maehara2021} \& N1 in \cite{Notsu2024}} \label{tb:prom_yzcmi_c1_two_spot}
\tabletypesize{\scriptsize}
\tablehead{
\colhead{Parameters} & \colhead{M1} & \colhead{N1}
}
\startdata
(Spot D) & & \\
Latitude (deg) & \multicolumn{2}{c}{$ -6.60 ^{+ 0.24 }_{- 0.19 }$} \\
Longitude (deg) & $ 38.73 ^{+ 0.17 }_{- 0.17 }$ & $ 76.26 ^{+ 0.17 }_{- 0.11 }$ \\
Radius (deg) & \multicolumn{2}{c}{$ 12.73 ^{+ 0.07 }_{- 0.08 }$} \\
Radius ($R_{\rm star}$) & \multicolumn{2}{c}{$ 0.222 ^{+ 0.001 }_{- 0.001 }$} \\
Radius ($10^{10}$ cm) & \multicolumn{2}{c}{$ 0.58 ^{+ 0.04 }_{- 0.10 }$} \\ \hline
(Spot E) & & \\
Latitude (deg) & \multicolumn{2}{c}{$ 41.17 ^{+ 0.30 }_{- 0.28 }$} \\
Longitude (deg) & $ -57.12 ^{+ 0.11 }_{- 0.17 }$ & $ -19.25 ^{+ 0.17 }_{- 0.11 }$ \\
Radius (deg) & \multicolumn{2}{c}{$ 15.92 ^{+ 0.01 }_{- 0.02 }$} \\
Radius ($R_{\rm star}$) & \multicolumn{2}{c}{$ 0.278 ^{+ 0.001 }_{- 0.001 }$} \\
Radius ($10^{10}$ cm) & \multicolumn{2}{c}{$ 0.72 ^{+ 0.05 }_{- 0.13 }$} \\
\enddata
\end{deluxetable*}

\clearpage
\bibliography{Main}{}

\begin{thebibliography}{}
\expandafter\ifx\csname natexlab\endcsname\relax\def\natexlab#1{#1}\fi
\providecommand{\url}[1]{\href{#1}{#1}}
\providecommand{\dodoi}[1]{doi:~\href{http://doi.org/#1}{\nolinkurl{#1}}}
\providecommand{\doeprint}[1]{\href{http://ascl.net/#1}{\nolinkurl{http://ascl.net/#1}}}
\providecommand{\doarXiv}[1]{\href{https://arxiv.org/abs/#1}{\nolinkurl{https://arxiv.org/abs/#1}}}

\bibitem[{Airapetian {et~al.}(2016)Airapetian, Glocer, Gronoff, H^^c3^^a9brard, \& Danchi}]{Airapetian2016}
Airapetian, V.~S., Glocer, A., Gronoff, G., H^^c3^^a9brard, E., \& Danchi, W. 2016, Nature Geoscience 2016 9:6, 9, 452, \dodoi{10.1038/ngeo2719}

\bibitem[{{Alvarado-G{\'o}mez} {et~al.}(2018){Alvarado-G{\'o}mez}, {Drake}, {Cohen}, {Moschou}, \& {Garraffo}}]{Alvarado-Gomez2018}
{Alvarado-G{\'o}mez}, J.~D., {Drake}, J.~J., {Cohen}, O., {Moschou}, S.~P., \& {Garraffo}, C. 2018, \apj, 862, 93, \dodoi{10.3847/1538-4357/aacb7f}

\bibitem[{{Astropy Collaboration} {et~al.}(2018){Astropy Collaboration}, {Price-Whelan}, {Sip{\H{o}}cz}, {G{\"u}nther}, {Lim}, {Crawford}, {Conseil}, {Shupe}, {Craig}, {Dencheva}, {Ginsburg}, {VanderPlas}, {Bradley}, {P{\'e}rez-Su{\'a}rez}, {de Val-Borro}, {Aldcroft}, {Cruz}, {Robitaille}, {Tollerud}, {Ardelean}, {Babej}, {Bach}, {Bachetti}, {Bakanov}, {Bamford}, {Barentsen}, {Barmby}, {Baumbach}, {Berry}, {Biscani}, {Boquien}, {Bostroem}, {Bouma}, {Brammer}, {Bray}, {Breytenbach}, {Buddelmeijer}, {Burke}, {Calderone}, {Cano Rodr{\'\i}guez}, {Cara}, {Cardoso}, {Cheedella}, {Copin}, {Corrales}, {Crichton}, {D'Avella}, {Deil}, {Depagne}, {Dietrich}, {Donath}, {Droettboom}, {Earl}, {Erben}, {Fabbro}, {Ferreira}, {Finethy}, {Fox}, {Garrison}, {Gibbons}, {Goldstein}, {Gommers}, {Greco}, {Greenfield}, {Groener}, {Grollier}, {Hagen}, {Hirst}, {Homeier}, {Horton}, {Hosseinzadeh}, {Hu}, {Hunkeler}, {Ivezi{\'c}}, {Jain}, {Jenness}, {Kanarek}, {Kendrew}, {Kern}, {Kerzendorf}, {Khvalko}, {King}, {Kirkby}, {Kulkarni},
  {Kumar}, {Lee}, {Lenz}, {Littlefair}, {Ma}, {Macleod}, {Mastropietro}, {McCully}, {Montagnac}, {Morris}, {Mueller}, {Mumford}, {Muna}, {Murphy}, {Nelson}, {Nguyen}, {Ninan}, {N{\"o}the}, {Ogaz}, {Oh}, {Parejko}, {Parley}, {Pascual}, {Patil}, {Patil}, {Plunkett}, {Prochaska}, {Rastogi}, {Reddy Janga}, {Sabater}, {Sakurikar}, {Seifert}, {Sherbert}, {Sherwood-Taylor}, {Shih}, {Sick}, {Silbiger}, {Singanamalla}, {Singer}, {Sladen}, {Sooley}, {Sornarajah}, {Streicher}, {Teuben}, {Thomas}, {Tremblay}, {Turner}, {Terr{\'o}n}, {van Kerkwijk}, {de la Vega}, {Watkins}, {Weaver}, {Whitmore}, {Woillez}, {Zabalza}, \& {Astropy Contributors}}]{Astropy2018}
{Astropy Collaboration}, {Price-Whelan}, A.~M., {Sip{\H{o}}cz}, B.~M., {et~al.} 2018, \aj, 156, 123, \dodoi{10.3847/1538-3881/aabc4f}

\bibitem[{{Bicz} {et~al.}(2022){Bicz}, {Falewicz}, {Pietras}, {Siarkowski}, \& {Pre{\'s}}}]{Bicz2022}
{Bicz}, K., {Falewicz}, R., {Pietras}, M., {Siarkowski}, M., \& {Pre{\'s}}, P. 2022, \apj, 935, 102, \dodoi{10.3847/1538-4357/ac7ab3}

\bibitem[{{Bisi} {et~al.}(2010){Bisi}, {Breen}, {Jackson}, {Fallows}, {Walsh}, {Miki{\'c}}, {Riley}, {Owen}, {Gonzalez-Esparza}, {Aguilar-Rodriguez}, {Morgan}, {Jensen}, {Wood}, {Owens}, {Tokumaru}, {Manoharan}, {Chashei}, {Giunta}, {Linker}, {Shishov}, {Tyul'bashev}, {Agalya}, {Glubokova}, {Hamilton}, {Fujiki}, {Hick}, {Clover}, \& {Pint{\'e}r}}]{Bisi2010}
{Bisi}, M.~M., {Breen}, A.~R., {Jackson}, B.~V., {et~al.} 2010, \solphys, 265, 49, \dodoi{10.1007/s11207-010-9602-8}

\bibitem[{{Cliver} {et~al.}(2022){Cliver}, {P{\"o}tzi}, \& {Veronig}}]{Cliver2022}
{Cliver}, E.~W., {P{\"o}tzi}, W., \& {Veronig}, A.~M. 2022, \apj, 938, 136, \dodoi{10.3847/1538-4357/ac847d}

\bibitem[{{Drake} {et~al.}(2016){Drake}, {Cohen}, {Garraffo}, \& {Kashyap}}]{Drake2016}
{Drake}, J.~J., {Cohen}, O., {Garraffo}, C., \& {Kashyap}, V. 2016, in IAU Symposium, Vol. 320, Solar and Stellar Flares and their Effects on Planets, ed. A.~G. {Kosovichev}, S.~L. {Hawley}, \& P.~{Heinzel}, 196--201, \dodoi{10.1017/S1743921316000260}

\bibitem[{{Feinstein} {et~al.}(2020){Feinstein}, {Montet}, {Ansdell}, {Nord}, {Bean}, {G{\"u}nther}, {Gully-Santiago}, \& {Schlieder}}]{Feinstein2020}
{Feinstein}, A.~D., {Montet}, B.~T., {Ansdell}, M., {et~al.} 2020, \aj, 160, 219, \dodoi{10.3847/1538-3881/abac0a}

\bibitem[{Fisher(1922)}]{Fisher1922}
Fisher, R.~A. 1922, Journal of the Royal Statistical Society, 85, 87.
\newblock \url{http://www.jstor.org/stable/2340521}

\bibitem[{{Gilbert} {et~al.}(2020){Gilbert}, {Barclay}, {Schlieder}, {Quintana}, {Hord}, {Kostov}, {Lopez}, {Rowe}, {Hoffman}, {Walkowicz}, {Silverstein}, {Rodriguez}, {Vanderburg}, {Suissa}, {Airapetian}, {Clement}, {Raymond}, {Mann}, {Kruse}, {Lissauer}, {Col{\'o}n}, {Kopparapu}, {Kreidberg}, {Zieba}, {Collins}, {Quinn}, {Howell}, {Ziegler}, {Vrijmoet}, {Adams}, {Arney}, {Boyd}, {Brande}, {Burke}, {Cacciapuoti}, {Chance}, {Christiansen}, {Covone}, {Daylan}, {Dineen}, {Dressing}, {Essack}, {Fauchez}, {Galgano}, {Howe}, {Kaltenegger}, {Kane}, {Lam}, {Lee}, {Lewis}, {Logsdon}, {Mandell}, {Monsue}, {Mullally}, {Mullally}, {Paudel}, {Pidhorodetska}, {Plavchan}, {Reyes}, {Rinehart}, {Rojas-Ayala}, {Smith}, {Stassun}, {Tenenbaum}, {Vega}, {Villanueva}, {Wolf}, {Youngblood}, {Ricker}, {Vanderspek}, {Latham}, {Seager}, {Winn}, {Jenkins}, {Bakos}, {Brice{\~n}o}, {Ciardi}, {Cloutier}, {Conti}, {Couperus}, {Di Sora}, {Eisner}, {Everett}, {Gan}, {Hartman}, {Henry}, {Isopi}, {Jao}, {Jensen}, {Law}, {Mallia}, {Matson},
  {Shappee}, {Le Wood}, \& {Winters}}]{Gilbert2020}
{Gilbert}, E.~A., {Barclay}, T., {Schlieder}, J.~E., {et~al.} 2020, \aj, 160, 116, \dodoi{10.3847/1538-3881/aba4b2}

\bibitem[{Gopalswamy(2016)}]{Gopalswamy2016}
Gopalswamy, N. 2016, Geoscience Letters 2016 3:1, 3, 1, \dodoi{10.1186/S40562-016-0039-2}

\bibitem[{{Hawley} {et~al.}(2014){Hawley}, {Davenport}, {Kowalski}, {Wisniewski}, {Hebb}, {Deitrick}, \& {Hilton}}]{Hawley2014}
{Hawley}, S.~L., {Davenport}, J. R.~A., {Kowalski}, A.~F., {et~al.} 2014, \apj, 797, 121, \dodoi{10.1088/0004-637X/797/2/121}

\bibitem[{{Houdebine} {et~al.}(1990){Houdebine}, {Foing}, \& {Rodono}}]{Houdebine1990}
{Houdebine}, E.~R., {Foing}, B.~H., \& {Rodono}, M. 1990, \aap, 238, 249

\bibitem[{{Ikuta} {et~al.}(2023){Ikuta}, {Namekata}, {Notsu}, {Maehara}, {Okamoto}, {Honda}, {Nogami}, \& {Shibata}}]{Ikuta2023}
{Ikuta}, K., {Namekata}, K., {Notsu}, Y., {et~al.} 2023, \apj, 948, 64, \dodoi{10.3847/1538-4357/acbd36}

\bibitem[{{Ikuta} {et~al.}(2020){Ikuta}, {Maehara}, {Notsu}, {Namekata}, {Kato}, {Notsu}, {Okamoto}, {Honda}, {Nogami}, \& {Shibata}}]{Ikuta2020}
{Ikuta}, K., {Maehara}, H., {Notsu}, Y., {et~al.} 2020, \apj, 902, 73, \dodoi{10.3847/1538-4357/abae5f}

\bibitem[{{Inoue} {et~al.}(2024){Inoue}, {Enoto}, {Namekata}, {Notsu}, {Honda}, {Maehara}, {Zhang}, {Lu}, {Uchida}, {Tsuru}, {Nogami}, \& {Shibata}}]{Inoue2024}
{Inoue}, S., {Enoto}, T., {Namekata}, K., {et~al.} 2024, \pasj, \dodoi{10.1093/pasj/psae001}

\bibitem[{{Kajikiya} {et~al.}(2025){Kajikiya}, {Namekata}, {Notsu}, {Maehara}, {Sato}, \& {Nogami}}]{Kajikiya2024}
{Kajikiya}, Y., {Namekata}, K., {Notsu}, Y., {et~al.} 2025, \apj, 979, 93, \dodoi{10.3847/1538-4357/ad91b9}

\bibitem[{{Kasting} {et~al.}(2014){Kasting}, {Kopparapu}, {Ramirez}, \& {Harman}}]{Kasting2014}
{Kasting}, J.~F., {Kopparapu}, R., {Ramirez}, R.~M., \& {Harman}, C.~E. 2014, Proceedings of the National Academy of Science, 111, 12641, \dodoi{10.1073/pnas.1309107110}

\bibitem[{{Kazachenko}(2023)}]{Kazachenko2023}
{Kazachenko}, M.~D. 2023, \apj, 958, 104, \dodoi{10.3847/1538-4357/ad004e}

\bibitem[{{Kipping}(2012)}]{Kipping2012}
{Kipping}, D.~M. 2012, \mnras, 427, 2487, \dodoi{10.1111/j.1365-2966.2012.22124.x}

\bibitem[{{Kobayashi} {et~al.}(2023){Kobayashi}, {Ise}, {Aoki}, {Kinoshita}, {Naito}, {Udo}, {Kunwar}, {Takahashi}, {Shibata}, {Mita}, {Fukuda}, {Oguri}, {Kawamura}, {Kebukawa}, \& {Airapetian}}]{Kobayashi2023}
{Kobayashi}, K., {Ise}, J.-i., {Aoki}, R., {et~al.} 2023, Life, 13, 1103, \dodoi{10.3390/life13051103}

\bibitem[{{Kowalski}(2024)}]{Kowalski2024}
{Kowalski}, A.~F. 2024, Living Reviews in Solar Physics, 21, 1, \dodoi{10.1007/s41116-024-00039-4}

\bibitem[{{Kretzschmar} {et~al.}(2010){Kretzschmar}, {de Wit}, {Schmutz}, {Mekaoui}, {Hochedez}, \& {Dewitte}}]{Kretzschmar2010}
{Kretzschmar}, M., {de Wit}, T.~D., {Schmutz}, W., {et~al.} 2010, Nature Physics, 6, 690, \dodoi{10.1038/nphys1741}

\bibitem[{{Kuhar} {et~al.}(2016){Kuhar}, {Krucker}, {Mart{\'\i}nez Oliveros}, {Battaglia}, {Kleint}, {Casadei}, \& {Hudson}}]{Kuhar2016}
{Kuhar}, M., {Krucker}, S., {Mart{\'\i}nez Oliveros}, J.~C., {et~al.} 2016, \apj, 816, 6, \dodoi{10.3847/0004-637X/816/1/6}

\bibitem[{{Kurita} {et~al.}(2020){Kurita}, {Kino}, {Iwamuro}, {Ohta}, {Nogami}, {Izumiura}, {Yoshida}, {Matsubayashi}, {Kuroda}, {Nakatani}, {Yamamoto}, {Tsutsui}, {Iribe}, {Jikuya}, {Ohtani}, {Shibata}, {Takahashi}, {Tokoro}, {Maihara}, \& {Nagata}}]{Kurita2020}
{Kurita}, M., {Kino}, M., {Iwamuro}, F., {et~al.} 2020, \pasj, 72, 48, \dodoi{10.1093/pasj/psaa036}

\bibitem[{Lammer {et~al.}(2007)Lammer, Lichtenegger, Kulikov, Grie^^c3^^9fmeier, Terada, Erkaev, Biernat, Khodachenko, Ribas, Penz, \& Selsis}]{Lammer2007}
Lammer, H., Lichtenegger, H.~I., Kulikov, Y.~N., {et~al.} 2007, Astrobiology, 7, 185, \dodoi{10.1089/ast.2006.0128}

\bibitem[{{Leitzinger} {et~al.}(2022){Leitzinger}, {Odert}, \& {Heinzel}}]{Leitzinger2022}
{Leitzinger}, M., {Odert}, P., \& {Heinzel}, P. 2022, \mnras, 513, 6058, \dodoi{10.1093/mnras/stac1284}

\bibitem[{{Lingam} {et~al.}(2018){Lingam}, {Dong}, {Fang}, {Jakosky}, \& {Loeb}}]{Lingam2018}
{Lingam}, M., {Dong}, C., {Fang}, X., {Jakosky}, B.~M., \& {Loeb}, A. 2018, \apj, 853, 10, \dodoi{10.3847/1538-4357/aa9fef}

\bibitem[{{Longcope}(2014)}]{Longcope2014}
{Longcope}, D.~W. 2014, \apj, 795, 10, \dodoi{10.1088/0004-637X/795/1/10}

\bibitem[{{Maehara} {et~al.}(2021){Maehara}, {Notsu}, {Namekata}, {Honda}, {Kowalski}, {Katoh}, {Ohshima}, {Iida}, {Oeda}, {Murata}, {Yamanaka}, {Takagi}, {Sasada}, {Akitaya}, {Ikuta}, {Okamoto}, {Nogami}, \& {Shibata}}]{Maehara2021}
{Maehara}, H., {Notsu}, Y., {Namekata}, K., {et~al.} 2021, \pasj, 73, 44, \dodoi{10.1093/pasj/psaa098}

\bibitem[{Morin {et~al.}(2008)Morin, Donati, Petit, Delfosse, Forveille, Albert, Auriere, Cabanac, Dintrans, Fares, Gastine, Jardine, Lignieres, Paletou, Velez, \& Theado}]{Morin2008}
Morin, J., Donati, J.~F., Petit, P., {et~al.} 2008, MNRAS, 390, 567, \dodoi{10.1111/j.1365-2966.2008.13809.x}

\bibitem[{{Moschou} {et~al.}(2019){Moschou}, {Drake}, {Cohen}, {Alvarado-G{\'o}mez}, {Garraffo}, \& {Fraschetti}}]{Moschou2019}
{Moschou}, S.-P., {Drake}, J.~J., {Cohen}, O., {et~al.} 2019, \apj, 877, 105, \dodoi{10.3847/1538-4357/ab1b37}

\bibitem[{{Namekata} {et~al.}(2017){Namekata}, {Sakaue}, {Watanabe}, {Asai}, {Maehara}, {Notsu}, {Notsu}, {Honda}, {Ishii}, {Ikuta}, {Nogami}, \& {Shibata}}]{Namekata2017}
{Namekata}, K., {Sakaue}, T., {Watanabe}, K., {et~al.} 2017, \apj, 851, 91, \dodoi{10.3847/1538-4357/aa9b34}

\bibitem[{Namekata {et~al.}(2022)Namekata, Maehara, Honda, Notsu, Okamoto, Takahashi, Takayama, Ohshima, Saito, Katoh, Tozuka, Murata, Ogawa, Niwano, Adachi, Oeda, Shiraishi, Isogai, Seki, Ishii, Ichimoto, Nogami, \& Shibata}]{Namekata2022}
Namekata, K., Maehara, H., Honda, S., {et~al.} 2022, Nature Astronomy, 6, 241, \dodoi{10.1038/s41550-021-01532-8}

\bibitem[{{Namekata} {et~al.}(2024{\natexlab{a}}){Namekata}, {Airapetian}, {Petit}, {Maehara}, {Ikuta}, {Inoue}, {Notsu}, {Paudel}, {Arzoumanian}, {Avramova-Boncheva}, {Gendreau}, {Jeffers}, {Marsden}, {Morin}, {Neiner}, {Vidotto}, \& {Shibata}}]{Namekata2024a}
{Namekata}, K., {Airapetian}, V.~S., {Petit}, P., {et~al.} 2024{\natexlab{a}}, \apj, 961, 23, \dodoi{10.3847/1538-4357/ad0b7c}

\bibitem[{{Namekata} {et~al.}(2024{\natexlab{b}}){Namekata}, {Ikuta}, {Petit}, {Airapetian}, {Vidotto}, {Heinzel}, {Wollmann}, {Maehara}, {Notsu}, {Inoue}, {Marsden}, {Morin}, {Jeffers}, {Neiner}, {Paudel}, {Avramova-Boncheva}, {Gendreau}, \& {Shibata}}]{Namekata2024b}
{Namekata}, K., {Ikuta}, K., {Petit}, P., {et~al.} 2024{\natexlab{b}}, \apj, 976, 255, \dodoi{10.3847/1538-4357/ad85df}

\bibitem[{{Namizaki} {et~al.}(2023){Namizaki}, {Namekata}, {Maehara}, {Notsu}, {Honda}, {Nogami}, \& {Shibata}}]{Namizaki2023}
{Namizaki}, K., {Namekata}, K., {Maehara}, H., {et~al.} 2023, \apj, 945, 61, \dodoi{10.3847/1538-4357/acb928}

\bibitem[{{Notsu} {et~al.}(2024){Notsu}, {Kowalski}, {Maehara}, {Namekata}, {Hamaguchi}, {Enoto}, {Tristan}, {Hawley}, {Davenport}, {Honda}, {Ikuta}, {Inoue}, {Namizaki}, {Nogami}, \& {Shibata}}]{Notsu2024}
{Notsu}, Y., {Kowalski}, A.~F., {Maehara}, H., {et~al.} 2024, \apj, 961, 189, \dodoi{10.3847/1538-4357/ad062f}

\bibitem[{{Nutzman} \& {Charbonneau}(2008)}]{Nutzman2008}
{Nutzman}, P., \& {Charbonneau}, D. 2008, \pasp, 120, 317, \dodoi{10.1086/533420}

\bibitem[{{Odert} {et~al.}(2020){Odert}, {Leitzinger}, {Guenther}, \& {Heinzel}}]{Odert2020}
{Odert}, P., {Leitzinger}, M., {Guenther}, E.~W., \& {Heinzel}, P. 2020, \mnras, 494, 3766, \dodoi{10.1093/mnras/staa1021}

\bibitem[{{Osten} \& {Wolk}(2015)}]{Osten2015}
{Osten}, R.~A., \& {Wolk}, S.~J. 2015, \apj, 809, 79, \dodoi{10.1088/0004-637X/809/1/79}

\bibitem[{Pearson(1900)}]{Pearson1900}
Pearson, K. 1900, The London, Edinburgh, and Dublin Philosophical Magazine and Journal of Science, 50, 157, \dodoi{10.1080/14786440009463897}

\bibitem[{{Ranjan} {et~al.}(2017){Ranjan}, {Wordsworth}, \& {Sasselov}}]{Ranjan2017}
{Ranjan}, S., {Wordsworth}, R., \& {Sasselov}, D.~D. 2017, \apj, 843, 110, \dodoi{10.3847/1538-4357/aa773e}

\bibitem[{{Ricker} {et~al.}(2015){Ricker}, {Winn}, {Vanderspek}, {Latham}, {Bakos}, {Bean}, {Berta-Thompson}, {Brown}, {Buchhave}, {Butler}, {Butler}, {Chaplin}, {Charbonneau}, {Christensen-Dalsgaard}, {Clampin}, {Deming}, {Doty}, {De Lee}, {Dressing}, {Dunham}, {Endl}, {Fressin}, {Ge}, {Henning}, {Holman}, {Howard}, {Ida}, {Jenkins}, {Jernigan}, {Johnson}, {Kaltenegger}, {Kawai}, {Kjeldsen}, {Laughlin}, {Levine}, {Lin}, {Lissauer}, {MacQueen}, {Marcy}, {McCullough}, {Morton}, {Narita}, {Paegert}, {Palle}, {Pepe}, {Pepper}, {Quirrenbach}, {Rinehart}, {Sasselov}, {Sato}, {Seager}, {Sozzetti}, {Stassun}, {Sullivan}, {Szentgyorgyi}, {Torres}, {Udry}, \& {Villasenor}}]{2015JATIS...1a4003R}
{Ricker}, G.~R., {Winn}, J.~N., {Vanderspek}, R., {et~al.} 2015, Journal of Astronomical Telescopes, Instruments, and Systems, 1, 014003, \dodoi{10.1117/1.JATIS.1.1.014003}

\bibitem[{{Science Software Branch at STScI}(2012)}]{Pyraf2012}
{Science Software Branch at STScI}. 2012, Pyraf: Python alternative for IRAF, Astrophysics Source Code Library.
\newblock \url{http://www.ascl.net}

\bibitem[{{Shibata} \& {Magara}(2011)}]{Shibata&Magara2011}
{Shibata}, K., \& {Magara}, T. 2011, Living Reviews in Solar Physics, 8, 6, \dodoi{10.12942/lrsp-2011-6}

\bibitem[{{Sun} {et~al.}(2022){Sun}, {T{\"o}r{\"o}k}, \& {DeRosa}}]{Sun2022}
{Sun}, X., {T{\"o}r{\"o}k}, T., \& {DeRosa}, M.~L. 2022, \mnras, 509, 5075, \dodoi{10.1093/mnras/stab3249}

\bibitem[{{Tian} {et~al.}(2015){Tian}, {Young}, {Reeves}, {Chen}, {Liu}, \& {McKillop}}]{Tian2015}
{Tian}, H., {Young}, P.~R., {Reeves}, K.~K., {et~al.} 2015, \apj, 811, 139, \dodoi{10.1088/0004-637X/811/2/139}

\bibitem[{{Tody}(1986)}]{IRAF1986}
{Tody}, D. 1986, in Society of Photo-Optical Instrumentation Engineers (SPIE) Conference Series, Vol. 627, Instrumentation in astronomy VI, ed. D.~L. {Crawford}, 733, \dodoi{10.1117/12.968154}

\bibitem[{{Vida} {et~al.}(2019){Vida}, {Leitzinger}, {Kriskovics}, {Seli}, {Odert}, {Kov{\'a}cs}, {Korhonen}, \& {van Driel-Gesztelyi}}]{Vida2019}
{Vida}, K., {Leitzinger}, M., {Kriskovics}, L., {et~al.} 2019, \aap, 623, A49, \dodoi{10.1051/0004-6361/201834264}

\bibitem[{{Vida} {et~al.}(2016){Vida}, {Kriskovics}, {Ol{\'a}h}, {Leitzinger}, {Odert}, {K{\H{o}}v{\'a}ri}, {Korhonen}, {Greimel}, {Robb}, {Cs{\'a}k}, \& {Kov{\'a}cs}}]{Vida2016}
{Vida}, K., {Kriskovics}, L., {Ol{\'a}h}, K., {et~al.} 2016, \aap, 590, A11, \dodoi{10.1051/0004-6361/201527925}

\bibitem[{{{\v{S}}vestka} {et~al.}(1962){{\v{S}}vestka}, {Kopeck{\'y}}, \& {Blaha}}]{Svestka1962}
{{\v{S}}vestka}, Z., {Kopeck{\'y}}, M., \& {Blaha}, M. 1962, Bulletin of the Astronomical Institutes of Czechoslovakia, 13, 37

\bibitem[{{Watanabe} {et~al.}(2017){Watanabe}, {Kitagawa}, \& {Masuda}}]{Watanabe2017}
{Watanabe}, K., {Kitagawa}, J., \& {Masuda}, S. 2017, \apj, 850, 204, \dodoi{10.3847/1538-4357/aa9659}

\bibitem[{{Watanabe} {et~al.}(2010){Watanabe}, {Krucker}, {Hudson}, {Shimizu}, {Masuda}, \& {Ichimoto}}]{Watanabe2010}
{Watanabe}, K., {Krucker}, S., {Hudson}, H., {et~al.} 2010, \apj, 715, 651, \dodoi{10.1088/0004-637X/715/1/651}

\bibitem[{{Watanabe} {et~al.}(2013){Watanabe}, {Shimizu}, {Masuda}, {Ichimoto}, \& {Ohno}}]{Watanabe2013}
{Watanabe}, K., {Shimizu}, T., {Masuda}, S., {Ichimoto}, K., \& {Ohno}, M. 2013, \apj, 776, 123, \dodoi{10.1088/0004-637X/776/2/123}

\bibitem[{{Yashiro} \& {Gopalswamy}(2009)}]{Yashiro&Gopalswamy2009}
{Yashiro}, S., \& {Gopalswamy}, N. 2009, in IAU Symposium, Vol. 257, Universal Heliophysical Processes, ed. N.~{Gopalswamy} \& D.~F. {Webb}, 233--243, \dodoi{10.1017/S1743921309029342}

\end{thebibliography}
\bibliographystyle{aasjournal}

\end{document}